\documentclass[conference]{IEEEtran}
\IEEEoverridecommandlockouts
\pdfoutput=1
\usepackage{cite}
\usepackage{amsmath,amssymb,amsfonts}
\usepackage{bm}
\usepackage{algorithmic}
\usepackage{graphicx}
\usepackage{textcomp}
\usepackage{subfigure}
\usepackage{xcolor}
\usepackage{multirow}
\usepackage[breaklinks,colorlinks,linkcolor=blue,citecolor=blue,urlcolor=blue]{hyperref}
\usepackage{makecell}
\usepackage{adjustbox}
\usepackage{booktabs}
\usepackage{tabularx}
\usepackage[linesnumbered,ruled]{algorithm2e}
\def\BibTeX{{\rm B\kern-.05em{\sc i\kern-.025em b}\kern-.08em
    T\kern-.1667em\lower.7ex\hbox{E}\kern-.125emX}}
\begin{document}
\newcommand{\mbm}{\bm}
\newcommand{\packmethod}{zero-aware greedy packing}
\newcommand{\tilemethod}{communication-aware operator tiling}
\newcommand{\ml}[1]{{\color{red}\bf [Meng: #1]}}
\newcommand{\xts}[1]{{\color{blue}\bf [Tianshi: #1]}}
\newcommand{\zwx}[1]{{\color{cyan}\bf [Zeng: #1]}}

\newtheorem{theorem}{Theorem}[section]
\newtheorem{example}{\textbf{Example}}[section]
\newtheorem{problem}{\textbf{Problem}}[section]

\newcommand{\method}{Falcon}
\newcommand{\offset}{\mathrm{offset}}
\newcommand{\Dec}{\mathrm{Dec}}
\newcommand{\Enc}{\mathrm{Enc}}
\newcommand{\Len}{\mathrm{Len}}

\title{Falcon: Accelerating Homomorphically Encrypted Convolutions for Efficient
Private Mobile \\ Network Inference}



\author{
\IEEEauthorblockN{Tianshi Xu$^{1,2}$, Meng Li$^{1,2*}$, Runsheng Wang$^{1,3,4}$, Ru Huang$^{1,3,4}$}
\IEEEauthorblockA{\textit{$^1$ School of Integrated Circuits, Peking University, Beijing, China}}
\IEEEauthorblockA{\textit{$^2$ Institute for Artificial Intelligence, Peking University, Beijing, China}}
\IEEEauthorblockA{\textit{$^3$ Institute of Electronic Design Automation, Peking University, wuxi, China}}
\IEEEauthorblockA{\textit{$^4$ Beijing Advanced Innovation Center for Integrated Circuits, Beijing, China}}
\IEEEauthorblockA{\textit{meng.li@pku.edu.cn}}
\thanks{This work was supported in part by the NSFC (62125401) and the 111 Project (B18001).}
}

\maketitle

\begin{abstract}
  Efficient networks, e.g., MobileNetV2, EfficientNet, etc, achieves
  state-of-the-art (SOTA) accuracy with lightweight computation. However,
  existing homomorphic encryption (HE)-based two-party computation (2PC) 
  frameworks are not optimized
  for these networks and suffer from a high inference overhead.
  We observe the inefficiency mainly comes from the packing algorithm, which 
  ignores the computation characteristics and the communication bottleneck of
  homomorphically encrypted depthwise convolutions.
  Therefore, in this paper, we propose \method, an effective dense packing
  algorithm for HE-based 2PC frameworks. \method~features a zero-aware greedy packing
  algorithm and a communication-aware operator tiling strategy to improve the
  packing density for depthwise convolutions. Compared to SOTA HE-based 2PC
  frameworks, e.g., CrypTFlow2, Iron and Cheetah, \method~achieves more than
  $15.6\times$, $5.1\times$ and $1.8\times$ latency
  reduction, respectively, at operator level. Meanwhile, at network level, \method~allows for
  $1.4$\% and $4.2$\% accuracy improvement over Cheetah on CIFAR-100 and Tiny
  Imagenet datasets with iso-communication, respecitvely.

\end{abstract}


\section{Introduction}


Recent years have witnessed the algorithmic breakthroughs as well as the 
compute and parameter explosion of convolutional neural networks (CNNs).
State-of-the-art (SOTA) efficient networks, e.g., MobileNetV2
\cite{sandler2018mobilenetv2}, EfficientNet \cite{tan2019efficientnet}, 
ConvNext \cite{liu2022convnet}, etc, achieve superhuman accuracy with
lightweight computation. Therefore, they are getting increasingly adopted in real-world
applications, including sensitive and private tasks such as face authentication
\cite{azouji2022efficientmask_face}, medical diagnosis \cite{kaissis2021end_medical}, etc. Privacy has thus emerged as a
major concern in the network deployment and there is a growing
demand for privacy-preserving deep neural network (DNN) inference \cite{Choi_Reagen_Wei_Brooks_Impala_2022,Demmler_Schneider_ABY_2015,Gupta_Kumaraswamy_Chandran_Gupta_LLAMA_2022,Kumar_Rathee_Chandran_Gupta_Rastogi_CrypTFlow_2020}.

Homomorphic encryption (HE)-based secure two-party computation (2PC) helps 
solve the following dilemma: the server, who owns a private model, and the client,
who owns private data, want to jointly apply the model to the data without
giving out the model or data.
HE-based 2PC not only enables the inference but also permits a cryptographically-strong privacy
protection, attracting a lot of attention in recent years \cite{huang2022cheetah,Dathathri_Saarikivi_Chen_Laine_Lauter_Maleki_Musuvathi_CHET_2019,Kim_Park_Kim_Kim_Ahn_HyPHEN_2023}.

As CNNs process high-dimensional tensors while HE computes over polynomials,
the first step of HE-based inference is the mapping from tensors 
to polynomials, whose coefficients are usually regarded as a one-dimensional vector,
denoted as packing \cite{Juvekar_Vaikuntanathan_gazelle_2018}. The packing algorithm directly
impacts the inference correctness and complexity and has been widely studied
to improve the HE-based 2PC efficiency for CNNs \cite{Juvekar_Vaikuntanathan_gazelle_2018,Mishra_Delphi_2020,rathee2020cryptflow2,huang2022cheetah,hao2022iron,Lou_Jiang_HEMET_2021}.

\begin{figure}[!tb]
  \includegraphics[width=\linewidth]{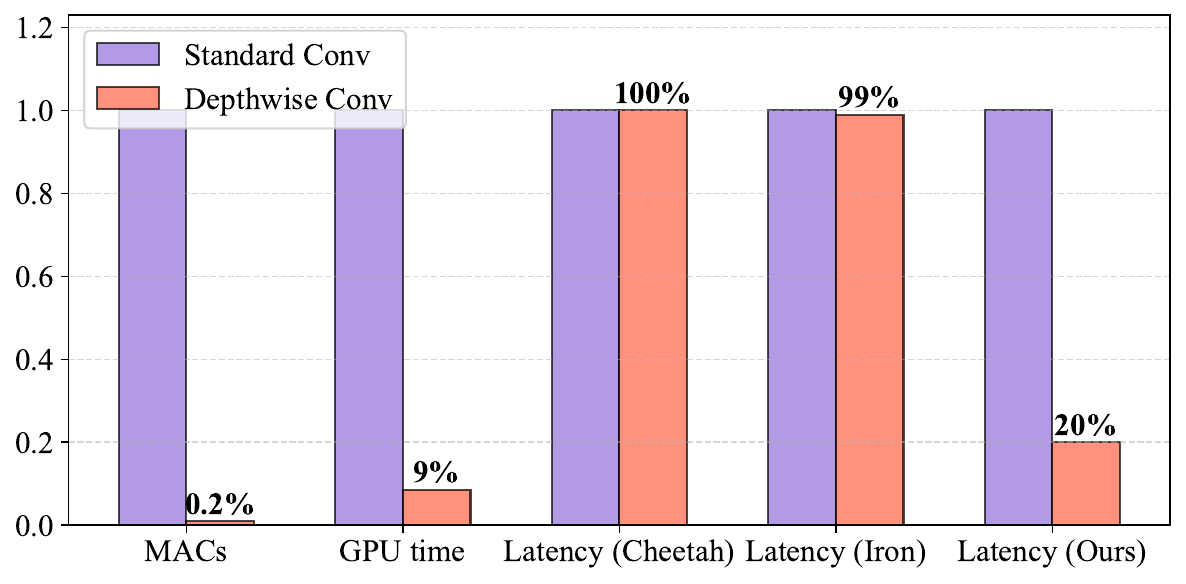}
  \caption{Compare a depthwise convolution with a standard convolution 
  on compute (measured by multiplication accumulations, i.e., MACs), 
  latency on GPU, as well as latency of two prior-art and our 2PC
  frameworks.
  }
  \label{fig:intro:dw_comp}
\end{figure}

However, existing HE-based 2PC frameworks are not optimized for
efficient networks, which
extensively use
depthwise and group convolutions to reduce the compute and latency
\cite{howard2017mobilenets,sandler2018mobilenetv2,tan2019efficientnet,tan2021efficientnetv2}.
As shown in Figure~\ref{fig:intro:dw_comp}, 
in two prior-art 2PC frameworks, the latency of a depthwise
convolution is almost the same as a standard convolution, indicating a high
inference inefficiency.

We observe the inefficiency comes from two limitations of the packing
algorithms in existing HE-based 2PC frameworks. On one hand, prior-art methods cannot leverage the
computation characteristics of depthwise convolutions. They have to pad the depthwise filter
with zeros and convert it back to a dense filter during packing \cite{rathee2020cryptflow2,huang2022cheetah}, leading to low packing density and high computation and communication overhead. 
On the other hand, they often ignore the communication to transfer
input and output ciphertexts are imbalanced and only focus on optimizing
the input communication \cite{Mishra_Delphi_2020,rathee2020cryptflow2,huang2022cheetah}.

In this paper, we propose \method, an effective dense packing algorithm for HE-based
2PC frameworks for privacy-preserving efficient network inference.
\method~features two main techniques to reduce inference latency for
depthwise and group convolutions. First, while zero padding is necessary
for computation correctness, \method~reuses the zero channels between padded
filters to improve the overall packing density. Secondly, we find the output
ciphertext transfer often dominates the overall communication and propose to
tile the input tensor and the depthwise convolution. Although the tiling
increases the input ciphertext transfer communication slightly, it drastically improves 
the output transfer efficiency. Our main contributions can be summarized as
follows:
\begin{itemize}
  \item We observe the bottleneck of existing HE-based 2PC frameworks when
    processing depthwise convolutions and propose \method~to improve the
    inference efficiency of SOTA lightweight networks.
  \item We propose a zero-aware greedy packing algorithm, which 
  formulates the
    depthwise filter packing as a shortest common superstring problem and
    proposes a greedy algorithm to improve the packing density.
  \item We propose a communication-aware operator tiling algorithm and achieve
    further communication reduction by balancing the input and output ciphertext
    transfer communication.
  \item \method~outperforms SOTA HE-based 2PC frameworks, including Cheetah and Iron, 
    with $1.4\sim5.4\times$ and $3.6\sim15.7\times$ communication reduction, 
    respectively, at operator level, especially for
    high polynomial orders. With iso-latency,
    \method~allows for $1.4$\% and $4.2$\% accuracy improvement on the CIFAR-100 and 
    Tiny Imagenet datasets over Cheetah, respectively.
\end{itemize}






\section{Preliminaries}


\subsection{Threat Model}

We focus on efficient privacy-preserving DNN inference involving two parites,
i.e., server and client, in which the server holds the proprietary DNN model
and the client owns private data \cite{huang2022cheetah,hao2022iron}. The model architecture, including
the number of layers as well as the types, dimensions, and bit widths for each
layer, is public to both the server and the client 
\cite{Kumar_Rathee_Chandran_Gupta_Rastogi_CrypTFlow_2020,Mishra_Delphi_2020,rathee2020cryptflow2}.
At the end of the protocol execution, the client learns the inference results 
without leaking any information to the server. Consistent with previous works
\cite{gilad2016cryptonets,hussain2021coinn,SecureML_2017,rathee2021sirnn}, we adopt an
\textit{honest-but-curious} security model in which both parties follow the
specification of the protocol but also try to learn more from the protocol
than allowed.

\subsection{Notations}\label{subsec:notation}

\begin{table}[!tb]
\centering
\caption{Notations used in the paper.}
\label{tab:notation}
\scalebox{0.85}{
\begin{tabular}{c|c}
\toprule
Notations & Meanings \\
\midrule
$[n]$ & $\{0,..,n-1\}$ for a non-negative integer $n$\\
\hline
$\lceil \cdot \rceil$,$\lfloor\cdot \rfloor$,$\lfloor\cdot \rceil$ & Ceiling, flooring, and rounding operations \\
\hline
$N,q$ & Polynomials degree, ciphertext bitwidth \\
\hline
$\hat{a}$ & \makecell{A polynomial and $\hat{a}[i]$ denotes the $i$-th \\
  coefficient of the polynomial} \\
\hline
$\lambda$ & Security parameter that measures the attack hardness \\
\hline
$\boxplus,\boxminus,\boxtimes $  & \makecell{Homomorphic addition, subtraction and multiplication} \\
\hline
$\Enc(\cdot), \Dec(\cdot)$ & Homomorphic encryption and decryption\\
\hline
$H, W, C$ & Height, width, and channels of convolution input tensor \\
\hline
$H^\prime, W^\prime, K$ & Height, width, and channels of convolution output tensor \\
\hline
$R, G, s$ & Convolution kernel size, group size, and stride \\
\bottomrule
\end{tabular}
}
\end{table}


Table~\ref{tab:notation} summarizes the notations used in this paper. 
The input, weight, and output tensors of a convolution operation are denoted
as $\bm{X}\in\mathbb{Z}^{C\times H\times W}$, 
$\bm{W}\in \mathbb{Z}^{K\times C\times R\times R}$, and $\bm{Y}\in\mathbb{Z}^{K\times H^\prime \times W^\prime }$,
where $\mathbb{Z}$ denotes the integer domain.
A convolution operation can be converted to a matrix-matrix multiplication 
$\bm{X}^\prime \bm{W}^\prime$ through the \texttt{im2col} operation \cite{chellapilla2006high}, 
where $\bm{X}^\prime\in\mathbb{Z}^{H^{\prime}W^{\prime} \times CR^2}$,
$\bm{W}^\prime \in \mathbb{Z}^{CR^2 \times K}$. For a group convolution,
$\bm{W}\in \mathbb{Z}^{K\times G\times R\times R}$. When $G=1$ and $K=C$,
it becomes a depthwise convolution. We also use $\mbm{X}_{i}$ to represent the $i$-th channel of
$\mbm{X}$, i.e., $\mbm{X}[i,:,:]$,
$\mbm{W}_{i,j}$ to represent the $j$-th channel of $i$-th filter, i.e., $\mbm{W}[i,j,:,:]$.
For simplicity, we also use $\mbm{W}_i$ denote $\mbm{W}_{i,i}$.

\subsection{HE-based 2PC Inference}

We review the basic flow of HE-based 2PC inference \cite{Liu_Juuti_MiniONN_2017,Mishra_Delphi_2020,Garimella_Ghodsi_Jha_Garg_Reagen_2022}. 
The 2PC framework
mainly involves two types of cryptographic primitives, including arithmetic
secret sharing (ASS) and HE. 
For ASS, an $l$-bit feature $x$ is shared by the server and the client 
additively as the sum of two values, say $\langle
x\rangle^S_{2^l}$ and  $\langle x\rangle^C_{2^l}$, and $x\equiv \langle
x\rangle^S_{2^l}+\langle x\rangle^C_{2^l}$\cite{SecureML_2017}. For simplicity,
we omit $l$ in the rest of the paper.
To compute the convolutions between the filter and the shared features,
HE is leveraged. HE enables to compute homomorphically on the ciphertext without
the need of decryption and thus, protects the data privacy. As in Table~\ref{tab:notation},
HE supports homomorphic addition, substruction, and multiplication.

\begin{figure}[!tb]
\centering
\includegraphics[width=1\linewidth]{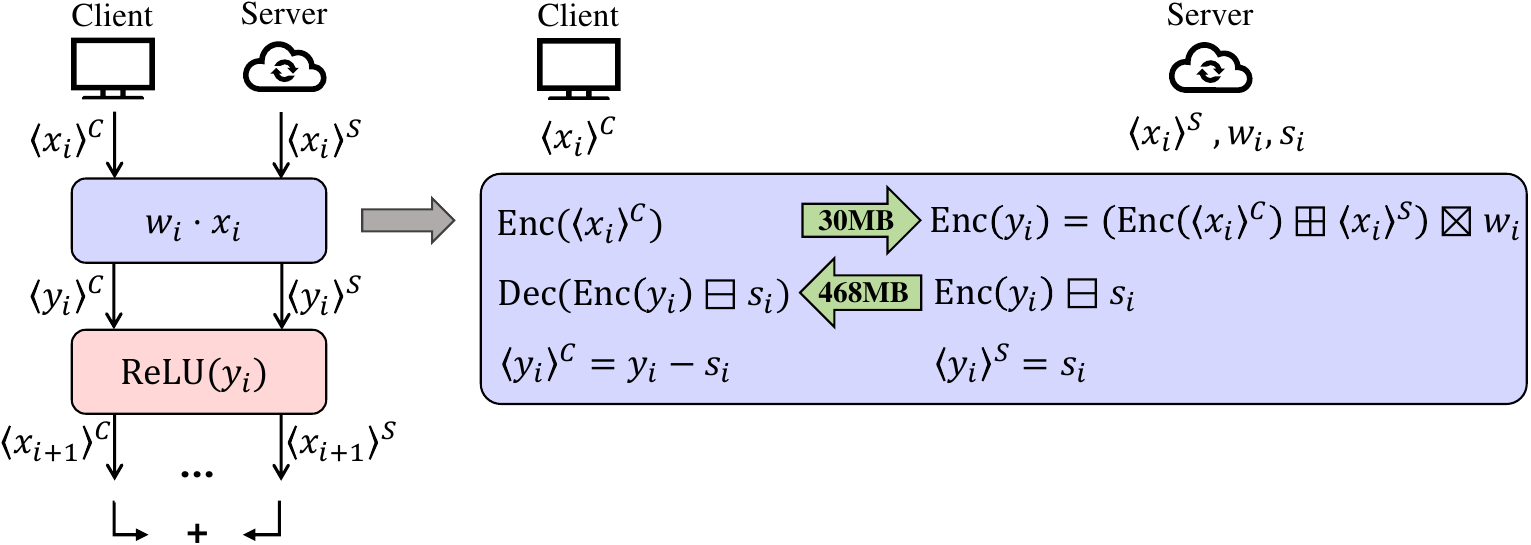}
\caption{Secure neural network inference based on HE (linear).}
\label{fig:secure_flow}
\end{figure}

In Figure \ref{fig:secure_flow}, the parameter of the $i$-th layer is represented as $w_i$, and 
the layer input and convolution output are denoted as $x_i$ and $y_i$, respectively.
During inference, the client and server hold additive shares denoted as $\langle \cdot\rangle^C,\langle \cdot\rangle^S$. The client encrypts its share as $\Enc(\langle x_i \rangle^C)$ and sends it to the server. The server perform homomorphic operations to obtain the output $\Enc(y_i) = (\Enc(\langle x_i \rangle^C) \boxplus \langle x_i \rangle^S) \boxtimes w_i$. Blinding of $y_i$ is done by subtracting a randomly sampled $s_i$ from $\Enc(y_i)$, and then, the result is sent to the client. The client decrypts it to obtain $\langle y_i \rangle^C=y_i - s_i$. The server outputs $s_i$ as $\langle y_i \rangle^S$. 
We omit the details of the ReLU protocols and refer interested readers to \cite{rathee2020cryptflow2,rathee2021sirnn,huang2022cheetah}.
Communication costs for a MobileNetV2 network performing secure inference on the CIFAR-100 dataset using Cheetah framework are highlighted as shown in Figure \ref{fig:secure_flow}: linear layers contribute $498\mathrm{MB}$ of communication in total, with $30\mathrm{MB}$ for input ciphertext transfer and $468\mathrm{MB}$ for output ciphertext transfer.

\subsection{Depthwise Convolution and Efficient Network}

\begin{figure}[!tb]
\centerline{\includegraphics[width=1\linewidth]{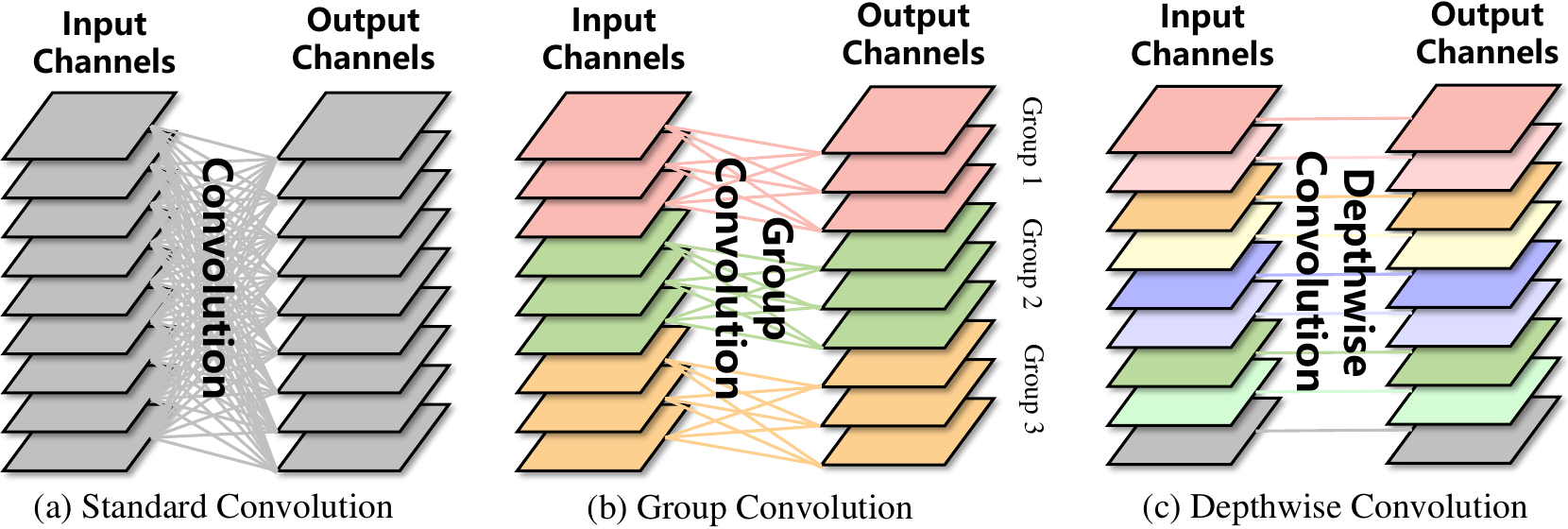}}
\caption{(a) Standard convolution, (b) group convolution and (c) depthwise convolution.}
\label{fig:depthwise_conv}
\end{figure}

As illustrated in Figure \ref{fig:depthwise_conv}, group convolution is a type
of sparsely connected convolution, which is first introduced in AlexNet \cite{krizhevsky2017imagenet_alexnet}. Standard convolution (Figure
\ref{fig:depthwise_conv}(a)) generates $K$ output features by applying
each filter over all $C$ input channels, resulting
in a computational cost of $O(KC)$. In contrast, group convolution (Figure
\ref{fig:depthwise_conv}(b)) reduces this computational cost by dividing the
input channels into $C/G$ mutually exclusive groups. Each group has $G$
channels and produces its own set of outputs, resulting in $O(KG)$
computational cost. When $G=1$, a group convolution becomes a depthwise convolution (Figure \ref{fig:depthwise_conv}(c)), where a single filter is applied to each
input channel.
Depthwise convolution is widely employed in efficient networks, including
MobileNetV1 \cite{howard2017mobilenets}, MobileNetV2 \cite{sandler2018mobilenetv2}, EfficientNet-Lite \cite{tan2019efficientnet}, etc.
These networks achieve SOTA accuracy 
with a high parameter and
compute efficiency.
For instance,
EfficientNet-Lite achieve similar accuracy compared to ResNet50 with $4 \times$ latency reduction \cite{tan2019efficientnet}.

\subsection{HE-based Convolution Protocols}
\label{subsec:HE based convolution protocols}


Since our approach primarily focuses on the HE-based convolution layer, we
first introduce how baseline methods, including CrypTFlow2 \cite{rathee2020cryptflow2},
Cheetah \cite{huang2022cheetah}, and Iron \cite{hao2022iron} perform a convolution
with a single input channel and single output channel (SISO).

\begin{figure}[!tb]
\centerline{\includegraphics[width=1\linewidth]{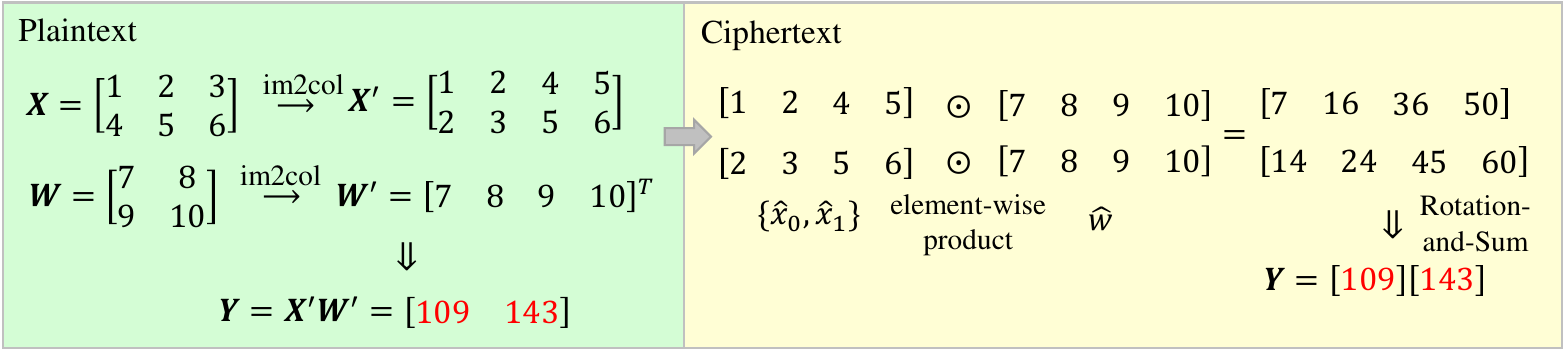}}
\caption{Illustration of SIMD packing: rotation is required to enable summation and leads to a high computation complexity.}
\label{fig:ex_simd}
\end{figure}

CrypTFlow2 leverages the SIMD 
packing to perform HE-based convolution,
which is also used by 
Gazelle \cite{Juvekar_Vaikuntanathan_gazelle_2018}, Delphi \cite{Mishra_Delphi_2020}, etc.
Figure \ref{fig:ex_simd} illustrates an example of using
SIMD packing to accomplish SISO convolution. 
It requires
\texttt{im2col} to transform a convolution into a matrix multiplication first, where
$\bm{X}^\prime\in\mathbb{Z}^{H^{\prime}W^{\prime} \times R^2}$ and
$\bm{W}^\prime\in\mathbb{Z}^{R^2 \times 1}$.
Then each row of $\bm{X}^\prime$ is encoded into a separate ciphertext vector, denoted as $\{\hat{x}_0, \ldots, \hat{x}_{H^{\prime} \times W^{\prime}-1}\}$. 
$\bm{W}^\prime$ is encoded as a plaintext vector $\hat{w}$.
With SIMD packing, a single homomorphic multiplication can realize the element-wise product between
the $\hat{x}_i$ and $\hat{w}$, resulting in $\hat{u}_i = \hat{x}_i \odot \hat{w}$.
To further sum all the elements in $\hat{u}_i$,
because $\hat{u}_i$ is encrypted, it can only be re-arranged by a rotation.
Hence, a ``rotation-and-sum'' step is needed before the final
results are calculated as in Figure~\ref{fig:ex_simd}.
Compared to a homomorphic multiplication, the computation complexity of a rotation is usually
much higher, leading to a high computation latency \cite{Juvekar_Vaikuntanathan_gazelle_2018,huang2022cheetah}.




\begin{figure}[!tb]
\centerline{\includegraphics[width=1\linewidth]{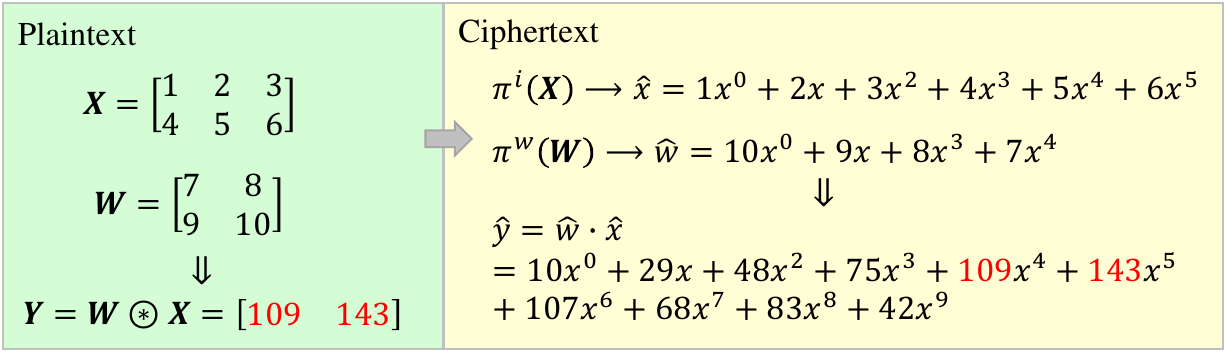}}
\caption{Polynomial coefficient packing proposed by Cheetah.}
\label{fig:ex_cheetah}
\end{figure}

\begin{figure}[!tb]
\centerline{\includegraphics[width=1\linewidth]{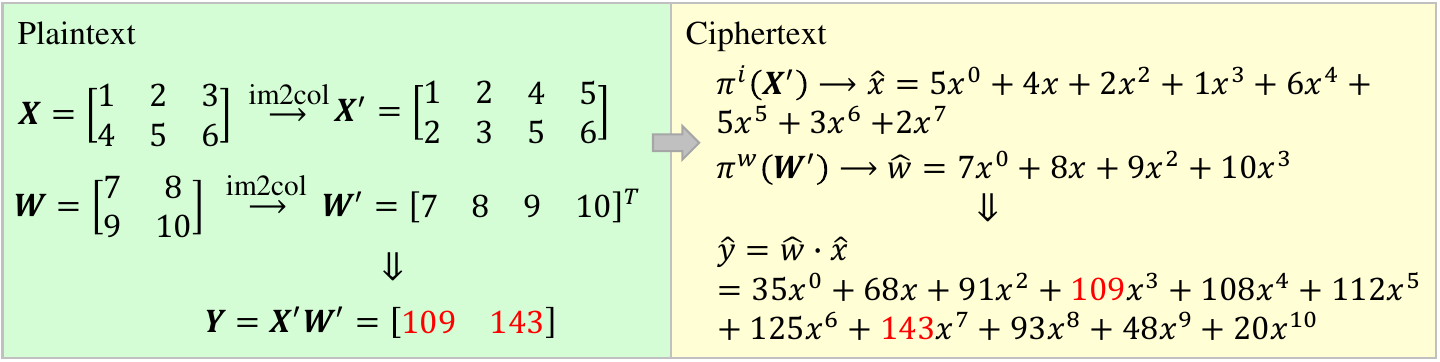}}
\caption{Polynomial coefficient packing proposed by Iron.}
\label{fig:ex_iron}
\end{figure}

To get rid of the expensive rotation, 
Cheetah \cite{huang2022cheetah} discovers that a polynomial multiplication already encompasses
convolution operations and proposes a polynomial coefficient encoding
method \cite{bian2021apas}. Specifically, Cheetah carefully designed two mappings:
$\pi^i$ and $\pi^w$ to convert the input and weight tensor to the polynomial coefficients,
i.e., one-dimensional vectors. 
The two polynomials are then
multiplied, and the correct convolution results can be directly extracted from
the polynomial coefficients as shown in Figure~\ref{fig:ex_cheetah}.
Iron \cite{hao2022iron} shares the same coefficient
packing algorithm as Cheetah and further optimizes the algorithm for matrix-matrix multiplications.
To support convolutions, Iron also requires running \texttt{im2col} first as shown in Figure~\ref{fig:ex_iron}.


\section{Inefficiency of Existing Solutions for Efficient Networks}
\label{sec:3}





From Figure~\ref{fig:intro:dw_comp}, we observe for prior-art 2PC frameworks, Cheetah and Iron,
the computation latency of a depthwise convolution is almost the same as that of a standard 
convolution, indicating a high inference overhead. We now analyze the origin of the overhead to
shed light on the directions of further optimization. Note that we do not discuss the SIMD packing
algorithm used in CrypTFlow2 as its rotation operations may
result in a latency gap of over 10$\times$ compared to Cheetah and Iron
following \cite{huang2022cheetah,hao2022iron} and as shown in the experimental
results.
Thus, it is not the focus of our analysis.

\subsection{Cost Analysis of HE-based Depthwise Convolution}
\label{subsec:metric}


The cost of a HE-based depthwise convolution primarily comes from computation
and communication between the client and the server. The computational complexity is
mainly determined by \textit{the number of homomorphic multiplications}, while the
communication consists of two parts. As in Figure
\ref{fig:secure_flow}, the first part is the transfer of input ciphertext
polynomials from the
client to the server, which equals to the product of the number of polynomials
and the communication of each polynomial.
The communication of a single polynomial scales with the polynomial degree and
the bit width of each coefficient, i.e., $\#Poly \times Nq$. The
second part is the transfer of output ciphertext polynomails from the server to
the client. 
Following the optimization of Cheetah,
the output transfer communication is $\#Poly \times (N+n)q$, 
where $n$ is the number of output elements in each polynomial.

\subsection{Inefficiency of Existing Packing Algorithms}
\label{subsec:Inefficiency of packing}


\begin{figure}[!tb]
\centering
\includegraphics[width=1\linewidth]{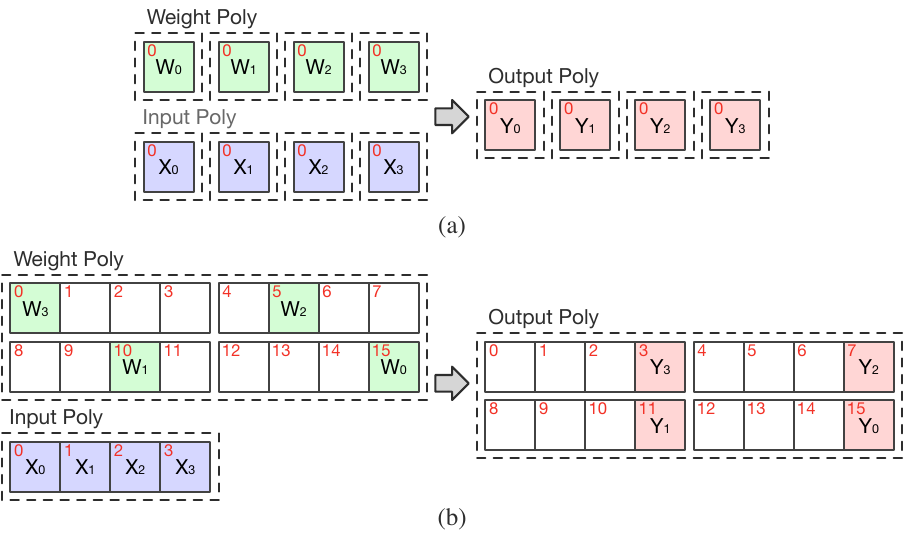}
\caption{Packing algorithm of (a) Iron and (b) Cheetah to support
  depthwise convolutions with multiple channels. Each square box represents a channel.
  Blank squares represent zero channels.
  Channels packed in the same polynomial are enclosed in the same dashed box.}
\label{fig:pack_cheetah_and_iron}
\end{figure}



In Section \ref{subsec:HE based convolution protocols}, we have already
introduced how Iron and Cheetah perform a single channel convolution and now, we 
focus on
analyzing their complexity for multi-channel cases. For ease of explanation, we
consider each channel as a basic unit and assume $HW \leq N$, i.e., each
channel convolution can be packed within a polynomial. When $HW > N$, we can follow
\cite{huang2022cheetah} to tile the input tensors.




\textbf{Iron} When extending to depthwise convolution with multiple channels,
Iron needs to perform \texttt{im2col} operation for each channel, transforming a single
depthwise convolution into $C$ matrix multiplications. Hence, Iron generates $C$
input polynomials, $C$ output polynomials, and in total $C$ homomorphic
multiplications. Following Section~\ref{subsec:metric}, the input transfer communication
becomes $C\times Nq$, while the output transfer communication is $C\times (N+H^\prime
W^\prime)q$.
Iron only compute one channel at one time, and each input polynomial and weight 
polynomial only packs $HW$ and $R^2$ elements, respectively, which demonstrates
a high packing redundancy.

\textbf{Cheetah} Cheetah implements the protocol for standard convolution, and
there are two approaches to extend Cheetah to support depthwise convolution.
The first approach is to execute the convolution separately for each channel,
which is essentially the same as Iron. A more
efficient approach is to restore the depthwise convolution to a standard
convolution by padding zero channels as shown in Figure~\ref{fig:pack_cheetah_and_iron}.
Then we can invoke Cheetah's standard convolution protocol, where the non-zero channels are
convolved with the input, while the
zero channels can be skipped directly. The advantage of this approach is that
it allows multiple channels to be packed into the same polynomial. 
Hence, for a depthwise convolution, Cheetah produces $\lceil \frac{CHW}{N} \rceil$ input polynomials and 
generates $C/\lceil \frac{N}{CHW}\rceil$ output polynomials, and requires $C/\lceil \frac{N}{CHW}\rceil$
homomorphic multiplications.
The communication for input
polynomials and output polynomials is $\lceil \frac{CHW}{N} \rceil Nq$
and $((C/\lceil \frac{N}{HWC}\rceil)N +CH^\prime W^\prime)q$, respectively.
Because $HW\le N$ and $\lceil \frac{N}{CHW}\rceil \ge 1$,
Cheetah's approach is strictly superior to Iron's approach in terms of both
computation and communication. However, in Cheetah's packing method, due
to the significant amount of zero channels introduced, $\frac{C-1}{C}$ of the
channels in weight polynomials are zero channels, leading to waste of compute 
and communication for output polynomials.

\subsection{Inefficiency of Tiling}



From the communication analysis in Section \ref{subsec:Inefficiency of
packing}, we observe that both Iron and Cheetah have a higher theoretical
communication cost for output polynomials compared to input polynomials. Our
experiments also confirm this observation as in Figure \ref{fig:secure_flow}.
However, both Iron and Cheetah's tiling strategies focus on optimizing the
communication cost of input polynomials: they try to pack as many input channels as
possible, resulting in more compact input ciphertext.
However, with the increase of the number of channels in the input polynomials,
Cheetah needs to add more zero channels, which reduces
the number of channels that can be packed into a weight polynomial. This
further increases the number of output polynomials, and raises the output
transfer communication.

\section{Depthwise Convolution Friendly Protocol}
\subsection{Overview}

We first provide an overview of our proposed method, \method.
We build \method~on top of Cheetah and address the significant overhead induced
by the padded zero channels by two steps, including zero-aware greedy packing
and communication-aware operator tiling.


\subsection{Zero-aware Greedy Packing}
\label{subsec:greedy packing}

\begin{figure}[!tb]
  \centering
  \includegraphics[width=1\linewidth]{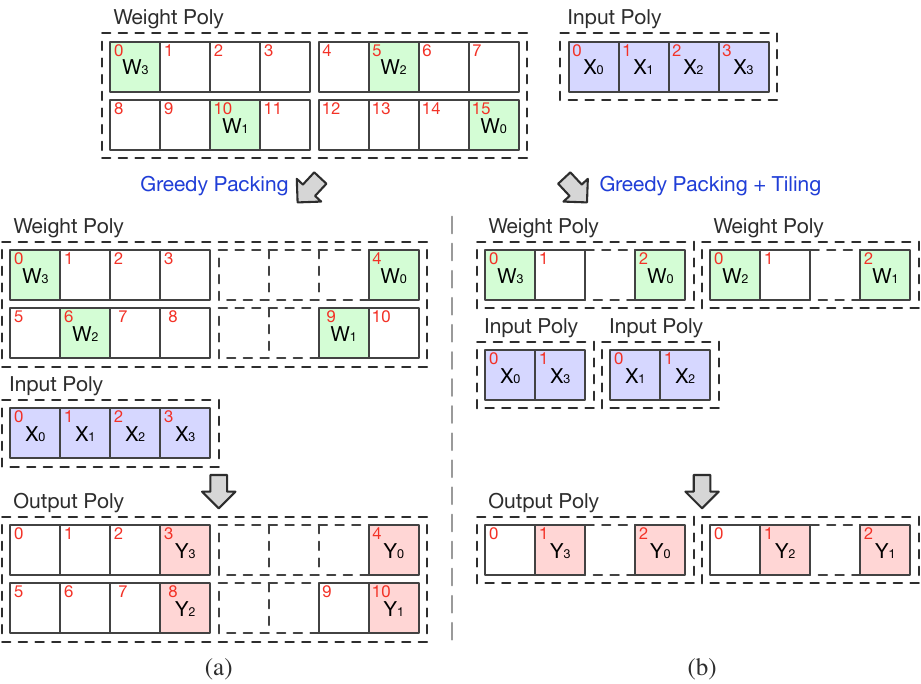}
  \caption{Illustration of (a) \packmethod~and (b) combining \packmethod~and \tilemethod. 
  Dashed boxes represent the zero channels reused by neighboring depthwise filters.}
\label{fig:example for gp and tile}  
\end{figure}

As mentioned in Section \ref{subsec:Inefficiency of packing}, the inefficiency
of Cheetah arises from a significant number of packed zero channels. 
As shown in Figure~\ref{fig:example for gp and tile}, for the depthwise convolution
with $4$ channels, after padding it to a standard convolution, the naive packing
algorithm in Cheetah maps the $4$ filters into
$\mbm{W}_{3}\mbm{000}\mbm{0}\mbm{W}_{2}\mbm{00}\mbm{00}\mbm{W}_{1}\mbm{0}\mbm{000}\mbm{W}_{0}$.
While the padded zero channels is necessary to eliminate the interference among 
channels to ensure the correctness of the computation, we make the following two
observations:
\begin{itemize}
    \item The computation order of different channels does not impact the computation correctness. For example,
        the 4 padded filters can be mapped as 
        $\mbm{W}_{3}\mbm{000}\mbm{000}\mbm{W}_{0}\mbm{0}\mbm{W}_{2}\mbm{00}\mbm{00}\mbm{W}_{1}\mbm{0}$ without
        impacting the computation of each filter.
    \item The zero channels of neighboring packed filters can be reused, e.g., $\mbm{W}_{3}\mbm{000}$ can share the 
        3 zero channels with $\mbm{000}\mbm{W}_{0}$, resulting in 
        $\mbm{W}_{3}\mbm{000}\mbm{W}_{0}\mbm{0}\mbm{W}_{2}\mbm{00}\mbm{W}_{1}\mbm{0}$.
\end{itemize}
The above two observations enable much denser packing to drastically reduce the number of padded zero channels.

More formally, let $\mbm{0}^{(k)}$ denote a sequence of $k$ consecutive zero channels, then,
the one-dimensional vector of $i$-th padded filter, denoted as $\tilde{\mbm{W}}_i$, can be represented as 
$\mbm{0}^{(C-1-i)} \mbm{W}_{i}\mbm{0}^{(i)}$, $\forall i\in[C - 1]$. 
Let $\tilde{\mbm{W}}$ denote the final mapping of the whole kernel, then, we observe as long as $\tilde{\mbm{W}}_i$
is a subsequence of $\tilde{\mbm{W}}$, the computation correctness can be guaranteed.
We omit the proof due to space constraints. Let $\Len(\tilde{\mbm{W}})$ denote the vector length of the final mapping,
then, we can formulate our zero-aware greedy packing as follows.

\begin{problem}
\label{problem:gp}
Given a set of vectors of padded filter $\{\tilde{\mbm{W}}_0, \ldots, \tilde{\mbm{W}}_{C-1} \}$, find the optimal vector $\tilde{\mbm{W}}$ that has the smallest length while preserving the computation correctness, i.e.,
\begin{align*}
    \mathrm{minimize} \quad & \Len(\tilde{\mbm{W}}) \\
    \mathrm{s.t.} \quad & \tilde{\mbm{W}}_i \text{ is a subsequence of } \tilde{\mbm{W}}, \forall i \in [C-1]
\end{align*}
\end{problem}

\begin{figure}[!tb]
\centerline{\includegraphics[width=1\linewidth]{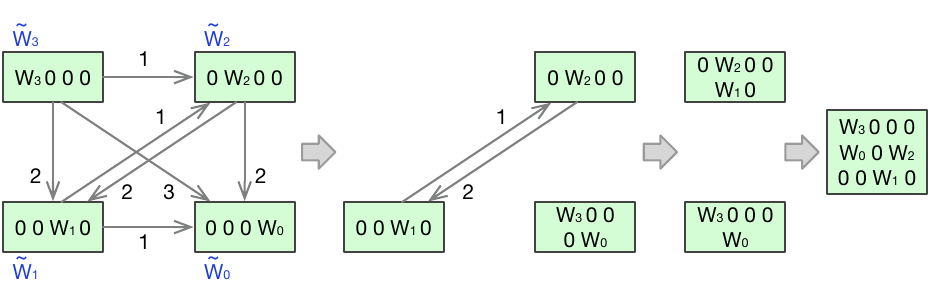}}
\caption{Illustration of greedy algorithm for SCS problem.}
\label{fig:alo_scs}
\end{figure}

Problem \ref{problem:gp} is equivalent to the shortest common superstring (SCS) problem,
which is proved to be NP-hard.
Lots of approximation algorithms have been developed to solve the problem.
We use Ukkonen's 1990 algorithm\cite{ukkonen1990linear_scs}. 
As shown in Figure \ref{fig:alo_scs}. First, we build a directed graph where vertices are $\{\tilde{\mbm{W}}_0,\tilde{\mbm{W}}_1,...,\tilde{\mbm{W}}_{C-1}\}$,
and the weight of the edge from $\tilde{\mbm{W}}_i$ to $\tilde{\mbm{W}}_j$ is the length of 
the longest suffix zero channels of $\tilde{\mbm{W}}_i$, which is also a prefix zero channels 
of $\tilde{\mbm{W}}_j$. Next, we randomly select one of the longest edges and merge the nodes
at both ends of the edge to form a new node and update the graph. Repeat this selecting and 
merging process until there are no more edges in the graph. Finally, merge the remaining nodes
together to obtain the results $\tilde{\mbm{W}}$.


\begin{figure}[!tb]
\centerline{\includegraphics[width=1\linewidth]{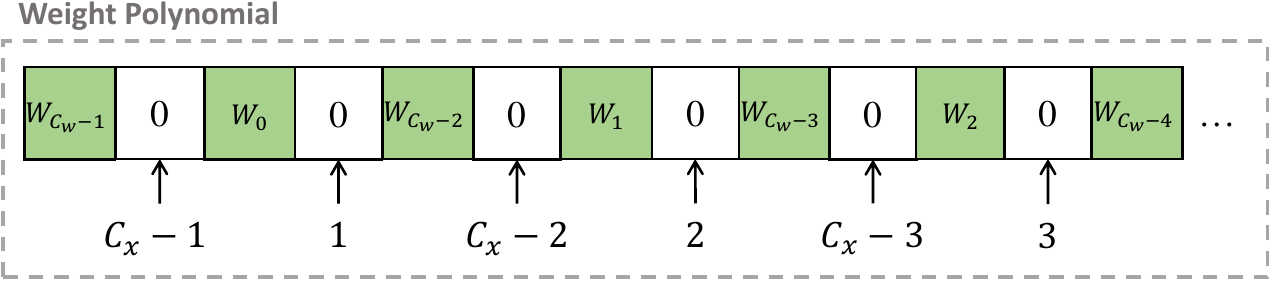}}
\caption{Illustration of how \packmethod~combines multiple channels.}
\label{fig:greedy packing}
\end{figure}

After applying the greedy algorithm, we discover that the solution always merges the $\tilde{\mbm{W}}_i$ and $\tilde{\mbm{W}}_{C-1-i}$
since they are symmetric in the location of zero channels. Furthermore, if the sequence starts with zeros, those zeros can 
be removed without affecting the corresponding coefficients in the output.
Based on the observation, we now formally derive the mapping function for 
our greedy packing algorithm described above.
Suppose an input polynomial packs $C_x$ channels, and a weight polynomial packs
$C_w$ channels. The sequence of weight polynomials is illustrated in Figure
\ref{fig:greedy packing}. Under the same $N$, we
have $C_x \ge C_w$ because additional zero channels need to be added for the
weight polynomials. Therefore, we let $C_x = kC_w$, where $k \in
\mathbb{Z^+}$, to prevent the channels corresponding to a weight polynomial
from being distributed across different input polynomials, which would result
in a doubling of the homomorphic multiplications. Now, we present the
coefficient mappings $\pi_{\mathrm{dwconv}}^i$ and $\pi_{\mathrm{dwconv}}^w$ to pack the three-dimensional input and weight tensors into one-dimensional vector. For simplicity, we require $\max (C_x\times
H\times W,C_w\times R\times R)\le N$, $C_x = C_w$ and $C_w$ is even.
\begin{align*}
& \hat{x}=\pi_{\mathrm{dwconv}}^i(\mathbf{X}) \text { st. } \hat{x}[c H W+i W+j]=\mathbf{X}[c,i,j] \\
    &\hat{w}=\pi_{\mathrm{dwconv}}^w(\mbm{W})\text { st. }\hat{w}[\offset(c^\prime)\times HW+O-l W-l^{\prime}] \notag\\
    &=\mbm{W}[c^\prime,l,l^{\prime}]
\end{align*}
where $O=W(h-1)+h-1$ and $\offset$ is: 
\begin{equation*}
    \offset(c^{\prime})=\left\{\begin{array}{l}
\left(C_x+2\right) \times\left(C_w-1-c^\prime\right), c^\prime \geqslant \frac{C_w}{2} \\
C_x+c^\prime\times\left(C_x+1\right), c^\prime<\frac{C_w}{2}
\end{array}\right.
\end{equation*}
The multiplication $\hat{y}=\hat{x}\cdot \hat{w}$ directly gives the 2-dimension depthwise convolution in some of the coefficients of the resulting polynomial: 
\begin{equation*}
    \mathbf{Y}\left[k^{\prime},i^{\prime}, j^{\prime}\right]= \hat{y}\left[(\offset(k^{\prime})+k^\prime)\times HW+O+i^{\prime} s W+j^{\prime} s\right]
\end{equation*}
where $\mathbf{Y}\in \mathbb{Z}^{C_x\times H^\prime \times W^\prime}$. 
$\offset(k^\prime)$ calculates the offset of weights at the
channel level, while $k^\prime$ calculates the offset of inputs at the channel
level. This is used to determine the starting position of each output channel,
while the encoding within a single channel remains consistent with Cheetah.

When $k \neq 1$, it means that the same input polynomial $\hat{x}$ needs to be
multiplied with multiple weight polynomials $\{\hat{w}_0, \hat{w}_1, ...,
\hat{w}_{k-1}\}$ to obtain the outputs $\{\hat{y}_0, \hat{y}_1, ...,
\hat{y}_{k-1}\}$. In order to perform packing in this case, we first
concatenate these weight polynomials end-to-end to form a larger polynomial
$\hat{w} = \hat{w}_0\hat{w}_1...\hat{w}_{k-1}$. Then, we apply packing to
$\hat{w}$, and $\hat{w}$ still satisfies $\hat{w}
= \pi_{\mathrm{dwconv}}^w(\mbm{W})$. Afterwards, $\hat{w}$ is divided into $k$
smaller polynomials, and the leading zeros of each polynomial are removed.

\textbf{Complexity Analysis}. Now we analyze the computational and
communication complexity of zero-aware greedy packing theoretically. We first
analyze the utilization of weight polynomial coefficients. Evidently, Cheetah's
utilization rate is $\frac{1}{C_x}$. Based on the analysis shown in
Figure \ref{fig:greedy packing}, our theoretical utilization is
$\frac{C_w}{C_x+1+(C_w/2-1)(C_x+2)}> \frac{2}{C_x+2}$. Considering the range of
values for $C_x$ in neural networks, zero-aware greedy packing can achieve
approximately twice the channel utilization rate compared to Cheetah. 

In terms of computational complexity, zero-aware greedy packing reduces the
number of polynomial multiplications by a factor of $\frac{2}{C_x+2}$ compared
to Cheetah. As for communication, zero-aware greedy packing reduces the
communication of the output polynomials from $(\lceil \frac{C_xHW}{N} \rceil
N + C_xH^\prime W^\prime)q$ to $(\lceil \frac{(C_x+2)HW}{2N} \rceil
N + C_xH^\prime W^\prime)q$ compared with Cheetah. In Section
\ref{subsec:Inefficiency of packing}, it has been proven that the computational
and communication complexity of Cheetah is strictly superior to Iron.
Therefore, our method outperforms both Cheetah and Iron in computation and communication.

\subsection{Communication-aware Operator Tiling}\label{subsec:tile}


As shown in Figure \ref{fig:secure_flow}, the communication of the output
polynomials is much higher than that of the input polynomials. This observation
aligns with the communication cost formula derived in Section
\ref{subsec:Inefficiency of packing}. Previous algorithms aim to
maximize the number of channels packed in one input polynomial
($C_x$)\cite{huang2022cheetah,hao2022iron,Juvekar_Vaikuntanathan_gazelle_2018,Mishra_Delphi_2020,rathee2020cryptflow2},
but this leads to reduced utilization of weight polynomial coefficients and
smaller $C_w$ because more zero channels need to be padded. Moreover, the
reduction in $C_w$ significantly increases the communication cost of
output polynomials, resulting in excessively high communication costs.

To address this issue, we propose a novel communication-aware operator tiling
method. Unlike previous methods, we conduct a theoretical analysis of
communication cost in depthwise convolution and formulate it as a nonlinear
programming problem. Our objective is to minimize communication while adhering
to polynomial coefficient encoding rules. Formally, the communication cost is
given by the equation:
\begin{equation*}
    \lceil \frac{C}{C_x}\rceil Nq+\lceil \frac{C}{C_w}\rceil(N+H^\prime W^\prime C_w)q
\end{equation*}
The first part of the equation represents the communication of the input
polynomials, where $\lceil \frac{C}{C_x}\rceil$ is the number of ciphertexts.
The second part corresponds to the output polynomials communication as
discussed in Section \ref{subsec:Inefficiency of packing}. By approximating the
ceiling function as $\lceil x\rceil \approx x$, we obtain:
\begin{equation*}
    NCq(\frac{1}{C_x}+\frac{1}{C_w})+H^\prime W^\prime Cq
\end{equation*}
Among them, only $\frac{1}{C_x}+\frac{1}{C_w}$ needs to be minimized, with all
other parameters known before executing the convolution protocol. We also have
two constraints: 1) polynomial coefficient encoding should not exceed the
degree $N$, and 2) $C_x$ must be a positive integer multiple of $C_w$.
Hence, we can formulate the communication optimization problem as follows:
\begin{align*}
\mathrm{minimize}\ \ \ \ \ &\ \  ||\frac{1}{C_x}+\frac{1}{C_w}|| \notag\\
\mathrm{s.t.}\ \ \ \ &(C_x+2)C_w\le\frac{2N}{HW}+2 \notag\\
&C_x=kC_w\text{ where }k \in \mathbb{Z^+}
\end{align*}
This is a solvable nonlinear programming problem with a relatively small
solution space. With $C_x\le \frac{N}{HW}$ and $C_x=kC_w$, the solution space
is at most $\frac{N}{HW} \times  \frac{N}{HW}$, allowing us to directly solve
it using a search algorithm with a complexity of $O((\frac{N}{HW})^2)$. An
example is illustrated in Figure \ref{fig:example for gp and tile} (b), after tiling, we reduce the size of weight polynomials and output polynomials.
\subsection{Extending to group convolution}

Furthermore, we discover that our method can be naturally extended to group
convolution. In group convolution, $C$ convolutional filters are divided into
$\frac{C}{G}$ groups, with each group containing $G$ convolutional filters,
and each filter consisting of $G$ channels. In this case, instead of treating
individual channels or filters as basic units, we consider a whole group of
convolutional filters (including $G$ kernels and $G^2$ channels) as a basic
unit. The packing between groups follows the rules of zero-aware greedy
packing, while within each group, it essentially becomes a standard convolution
and can be directly packed using the Cheetah approach for standard convolutions.

\section{Experimental Results}

\subsection{Experimental Setup}
\method~is built on top of the SEAL library \cite{sealcrypto}, the EMP
toolkit \cite{emp-toolkit} and OpenCheetah\cite{huang2022cheetah} in C++. We also use the Ezpc
library\cite{chandran2017ezpc} to evaluate
CrypTFlow2\cite{rathee2020cryptflow2}. 
Consistent with
\cite{Shen_Dong_Fang_Shi_Wang_Pan_Shi_ABNN2_2022,SecureML_2017}, we simulate
a WAN network setting via Linux Traffic Control, where the bandwidth is 9 MBps.
All the following experiments are performed on machines with 2.2 GHz Intel Xeon
CPU and 256GB RAM. Following \cite{hao2022iron,huang2022cheetah}, we set
$N=4096$ for most of the cases.

\subsection{Microbenchmark Evaluation}


\paragraph{Single Depthwise Convolution Evalution}

\begin{figure}[!tb]
\centering
\includegraphics[width=1\linewidth]{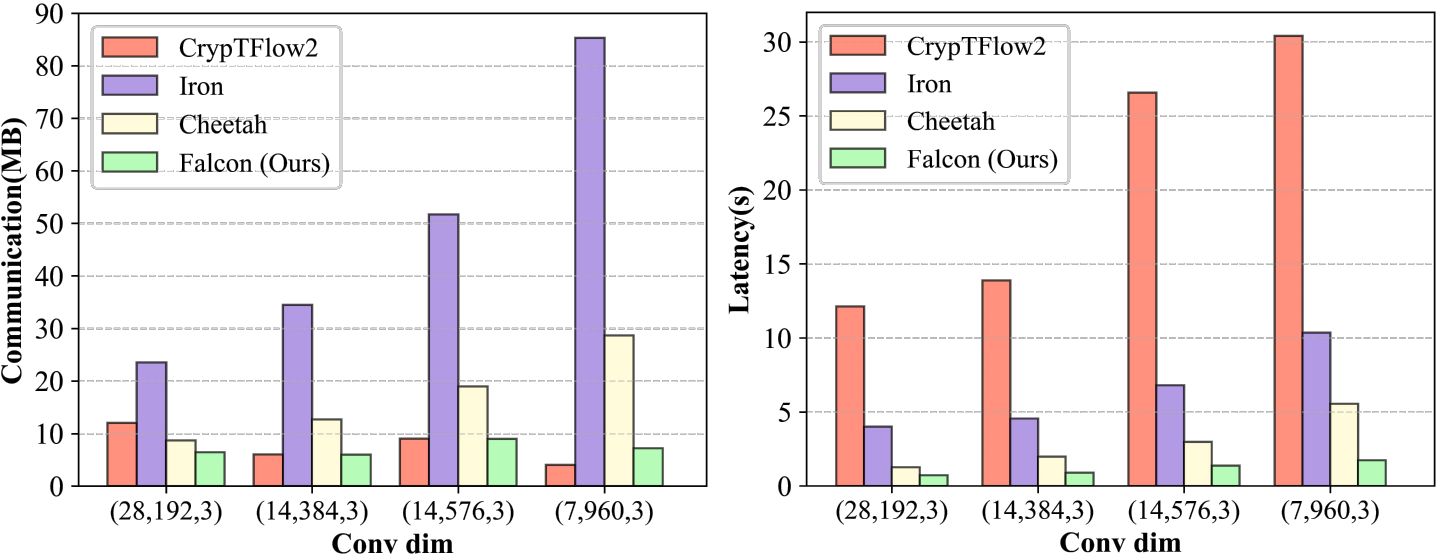}
  \caption{(a) Communication and (b) latency comparison of depthwise
  convolutions of different dimensions (the tuple represents the input feature 
  height, channels, and kernel size).}
\label{fig:dwconv_eval}
\end{figure}

We compare the performance of \method~with CrypTFlow2, Iron, and Cheetah for
a single depthwise convolution. As shown in Figure \ref{fig:dwconv_eval},
\method~achieves a communication reduction of $1.4\sim4.0\times$ and a latency
reduction of $1.8\sim3.2\times$ compared to 
Cheetah. Compared to Iron, \method~achieves a communication reduction of
$3.6\sim11.8\times$. It is important to note that CrypTFlow2 (with SIMD) has
relatively low communication. However, as mentioned in Section \ref{sec:3}, the
SIMD method has high computational complexity, resulting in significant
latency. In comparison, our method achieves a latency reduction of $15.6\sim19.9\times$.

\paragraph{Depthwise Convolution under Different Polynomial Degree}

\begin{figure}[!tb]
  \centering
  \includegraphics[width=1\linewidth]{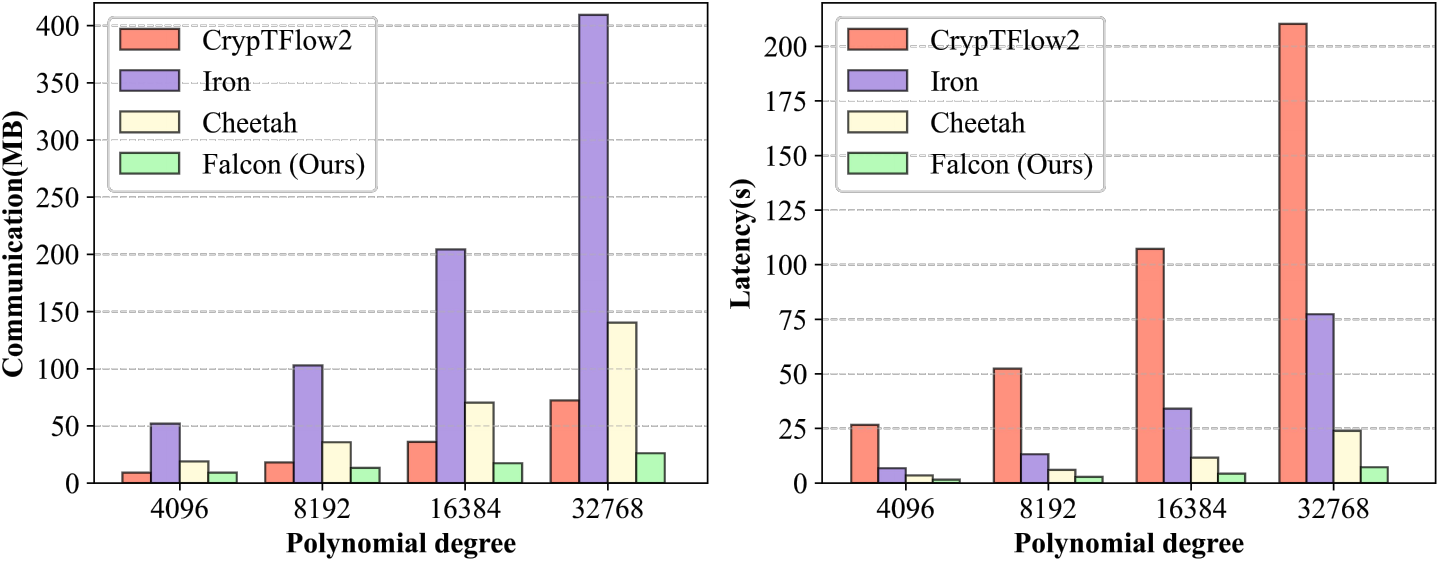}
  \caption{(a) Communication and (b) latency comparison of a depthwise
  convolution under different polynomial degrees.}
  \label{fig:n_diff_eval}
\end{figure}

We benchmark \method~against CrypTFlow2, Cheetah, and Iron, 
for different polynomial degrees. We select
a depthwise convolution with input dimension of $(14, 576, 3)$ for evaluation. 
As shown in Figure~\ref{fig:n_diff_eval}, overall, with the increase of $N$,
\method~achieves a higher speedup compared to existing methods due to a larger
packing optimization space. Compared to Cheetah, we
achieve a communication reduction of $4.0\times$ at $N=4096$ and an even
greater reduction of $11.1\times$ at $N=32768$. Compared to Iron, our
method achieves a remarkable communication reduction of $32.9\times$ when
$N=32768$.


\paragraph{Depthwise Convolution under Different Bandwidth} 

We also benchmark the latency of \method~against CrypTFlow2,
Cheetah, and Iron under different bandwidth settings: $9\mathrm{MBps}$,
$44\mathrm{MBps}$, and $384\mathrm{MBps}$, representing WAN settings in
\cite{Shen_Dong_Fang_Shi_Wang_Pan_Shi_ABNN2_2022,SecureML_2017}, WAN and LAN
settings in Cheetah\cite{huang2022cheetah}, respectively. In Table
\ref{table:eval_diff_bandwidth}, our methods achieve a significant latency
reduction of approximately $5.6\sim7.35 \times$  compared to Cheetah, and this
advantage increases as the bandwidth decreases. We also observe that CrypTFlow2
and Iron are not sensitive to bandwidth. For CrypTFlow2, this is because it
suffers from a high computational complexity while for Iron, we hypothesize the
latency bottleneck comes from the frequent encryption and decryption.

\begin{table}[!tb]
  \caption{Latency (s) comparison under different bandwidths.}
\label{table:eval_diff_bandwidth}
\label{tab:impact_bw}
\centering
\begin{tabular}{|c|c|c|c|c|c|}
\hline
\multirow{2}{*}{\textbf{Bandwidth}}&\multicolumn{4}{c|}{\textbf{Methods}} \\
\cline{2-5}
 & \textbf{CrypTFlow2}& \textbf{Iron}& \textbf{Cheetah}& \textbf{\method~(ours)} \\
\hline
  384MB/s \cite{huang2022cheetah} &29.66 &10.23&2.44&0.43\\
\hline
44MB/s \cite{huang2022cheetah} & 30.24&10.49&3.04&0.45\\
\hline
9MB/s\cite{Shen_Dong_Fang_Shi_Wang_Pan_Shi_ABNN2_2022,SecureML_2017} & 30.86&10.68&5.56&0.76\\
\hline
\end{tabular}
\end{table}

\paragraph{Group Convolution Evaluation} 

We also evaluate the performance of \method~for group
convolutions. We select a depthwise convolution with dimension of $(14,576,3)$ and
benchmark its performance under different group sizes. As shown in
Table~\ref{table:eval_group_conv}, our method consistently achieves better communication
for different group sizes. Additionally, we notice that, except for Iron,
the other methods are not sensitive to the group size. 
The main reason is that
both CrypTFlow2 and Cheetah expand group convolutions into standard
convolutions. In contrast, Iron's approach leads to an increase in matrix
multiplication dimension as the group size grows. When the dimension exceed
the polynomial degree $N$, it results in a significant increase of communication.

\begin{table}[!tb]
\caption{Communication (MB) comparison under different group sizes.}
\label{table:eval_group_conv}
\begin{center}
\begin{tabular}{|c|c|c|c|c|c|}
\hline
\multirow{2}{*}{\textbf{Group size}}&\multicolumn{4}{c|}{\textbf{Methods}} \\
\cline{2-5}
 & \textbf{CrypTFlow2}& \textbf{Iron}& \textbf{Cheetah}&  \textbf{\method~(ours)} \\
\hline
1 &9.00&51.70&18.94&8.98\\
\hline
2&9.00&51.70&18.94&8.98\\
\hline
4&9.00&68.44&18.94&8.98\\
\hline
8&9.00&102.73&18.94&8.98\\
\hline
\end{tabular}
\label{tab:impact_gw}
\end{center}
\end{table}

\begin{table*}[!tb]
\centering
\caption{End-to-end comparison with prior-art works
  on different architectures and datasets (the inference complexity on Cifar10
  and Cifar100 is essentially the same).}
  \label{table:eval_endtoend}
\begin{tabular}{|c|c|c|c|c|c|c|c|}
\hline
\multirow{2}{*}{\textbf{Network}}&\multirow{2}{*}{\textbf{Dataset}}
 & \multicolumn{2}{c|}{\textbf{CrypTFlow2}}&
  \multicolumn{2}{c|}{\textbf{Cheetah}}&
  \multicolumn{2}{c|}{\textbf{\method~(Ours)}}\\
\cline{3-8}
& & Comm. (GB)  & Latency (s) &  Comm. (GB)  & Latency (s) & Comm. (GB)  & Latency (s)\\
\hline
MobileNetV2&Cifar& 2.32&622.43&0.60&132.03&\textbf{0.46}&\textbf{109.27}\\
\hline
MobileNetV2& Tiny Imagenet& 2.42&1067.34&0.70&168.35&\textbf{0.54}&\textbf{130.41}\\
\hline
EfficientNet-lite& Cifar &0.83&563.90&0.50&103.48&\textbf{0.34}&\textbf{76.61}\\
\hline
EfficientNet-lite& Tiny Imagenet& 1.28&915.36&0.64&146.37&\textbf{0.45}&\textbf{107.44}\\
\hline
\end{tabular}
\end{table*}

\begin{figure*}[!tb]
\centerline{\includegraphics[width=0.95\linewidth]{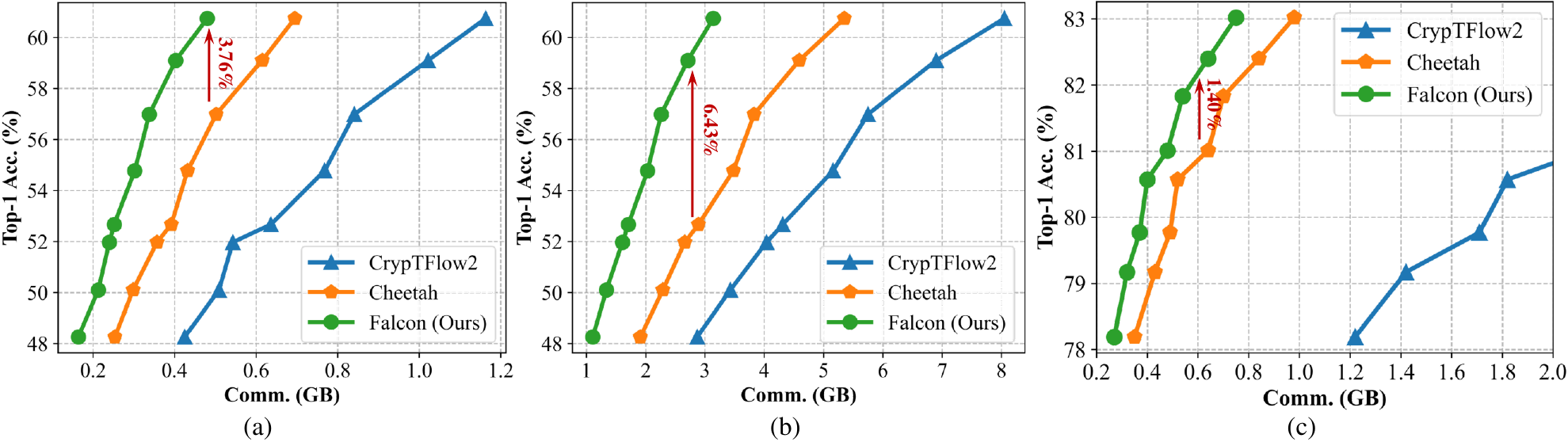}}
\caption{Accuracy comparison with Cheetah and CrypTFlow2 on (a) Tiny Imagenet, N=4096; (b) Tiny Imagenet, N=32768; (c) Cifar-100, N=4096.}
\label{fig:eval_acc}
\end{figure*}

\begin{figure}[!tbp]
\centerline{\includegraphics[width=1\linewidth]{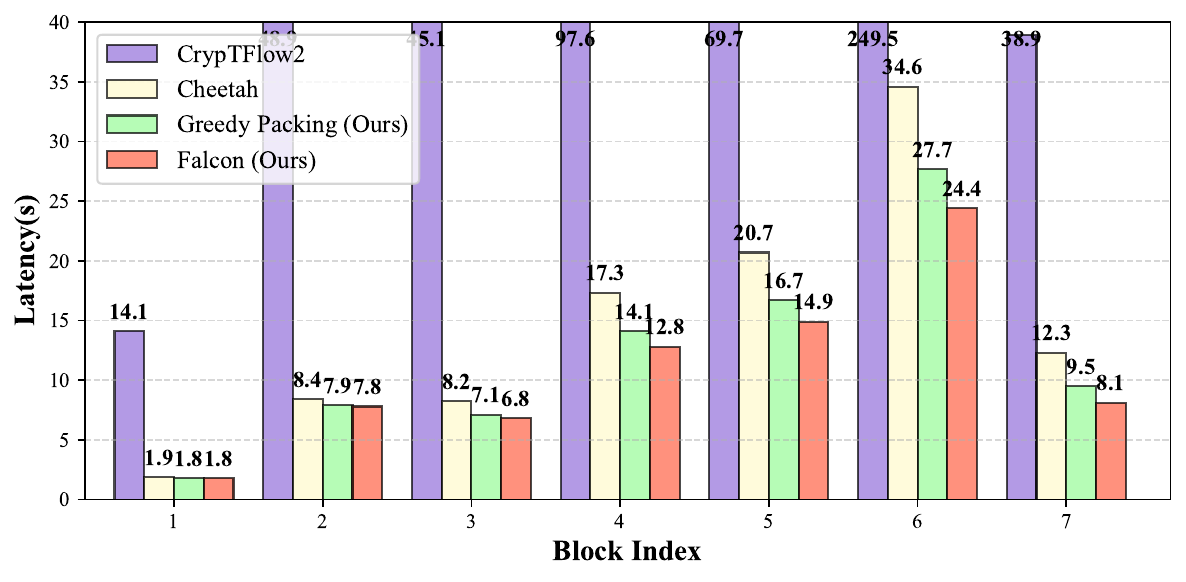}}z
\caption{Block-wise comparison of latency on MobileNetV2/Tinyimagenet, N=4096.}
\label{fig:block_wise}
\end{figure}

\subsection{End-to-end Inference Evaluation}
\paragraph{Communication and Latency Comparison}


We now perform an end-to-end inference evaluation compared to prior-art
methods, including CrypTFlow2 and Cheetah. Due to the unavailability of
open-source code for Iron, we find it difficult to reproduce all of its linear
and nonlinear protocols. As a result, we are unable to evaluate its performance
in the end-to-end inference. 
We apply \method~to two widely used efficient networks,
MobileNetV2 and EfficientNet-lite, and evaluate their performance on the
Cifar10, Cifar100 \cite{krizhevsky2009learning_cifar}, and Tiny Imagenet \cite{deng2009imagenet} datasets.
We set the bandwidth to $9\mathrm{MBps}$ and $N=4096$.

\textbf{Results and analysis}. As shown in Table \ref{table:eval_endtoend},
compared to Cheetah, \method~reduces the communication by $1.31\times$ and the
latency by more than $1.2\times$ on the Cifar10 and Cifar100 datasets.
On the Tiny Imagenet dataset, the advantage of \method~is greater, with
a communication reduction of $1.44\sim1.48\times$ and a latency reduction
of $1.35\sim1.36\times$. 
These experimental results resonate with the theoretical
expectations. 
For Tiny Imagenet, the networks have more layers which have
smaller dimensions $H$ and $W$ but larger channel size $C$, resulting in
a larger packing optimization space.


\paragraph{Networks Accuracy Comparison}

We also benchmark the accuracy of MobileNetV2 networks with different
channel-widths on Cifar100 and Tiny Imagenet for CrypTFlow2,
Cheetah, and \method~(Ours). Additionally, we selecte $N=4096$ and $N=32768$ to
demonstrate the superiority of our method with larger $N$.

\textbf{Results and analysis}. As shown in Figure~\ref{fig:eval_acc},
\method~consistently outperforms the other methods across different network
widths. In Tiny Imagenet,when $N=4096$, with approximately $0.5\mathrm{GB}$ of communication
costs, \method~achieves $60.75\%$ top-1 accuracy, surpassing Cheetah and
CrypTFlow2 by $3.76\%$ and $10.65\%$, respectively. When $N$ is
further increased to $32768$, \method~maintains a substantial lead in
top-1 accuracy over Cheetah and CrypTFlow2. With approximately $2.8\mathrm{GB}$
of communication, \method~outperforms them by $6.43\%$ and $10.81\%$,
respectively. For the Cifar100 dataset, as the network exhibits strong capability on it, increasing the width does not bring significant accuracy improvement. However, Falcon still achieves an average accuracy improvement of $1.4$\%.


\paragraph{Block-wise Latency Comparison}

We show the block-wise latency comparison in Figure \ref{fig:block_wise}. As we
can observe, \method~achieves more latency reduction in deeper layers of the network, as
$H,W$ becomes smaller and $C $ becomes larger, resulting in a larger packing
optimization space. Additionally, \method~reduces the latency for all the blocks,
highlighting the effectiveness of \method.

\section{Conclusion}
\label{sec:conclu}

In this paper, we propose \method, a dense packing algorithm for homomorphically encrypted 
depthwise and group convolution for HE-based 2PC network inference. \method~leverages the
computation characteristics of the depthwise convolution and directly optimize the communication
bottleneck. \method~features a zero-aware greedy packing algorithm and a communication-aware operator
tiling strategy to significantly reduce the communication for efficient networks. Compared with
prior-art works, e.g., Cheetah and Iron, \method~achieves a latency reduction of $1.8\sim 3.2\times$
at the operator level and $1.21\sim 1.36\times$ at the network level, which
translates to 1.4\% and 4.2\% accuracy improvement with iso-communication. These 
improvements enhance the practicality of HE-based 2PC inference for efficient networks.


\newpage

\bibliographystyle{IEEEtran}
\bibliography{bibs/ref.bib}

\end{document}